\documentclass[aps,twocolumn,superscriptaddress,showpacs,preprintnumbers,amsmath,amssymb,amsfonts,nofootinbib]{revtex4}

\usepackage{graphicx}
\usepackage{bbm}

\usepackage{color}

\def\0#1#2{\frac{#1}{#2}}
\newcommand{\id}{\mathbbm{1}}
\def\s0#1#2{\mbox{\small{$ \frac{#1}{#2} $}}}

\def\CP{{\mathcal P}}
\def\CC{{\mathcal C}}

\def\CO{{\mathcal O}}

\newcommand{\Eqref}[1]{Eq.~\eqref{#1}}

\newcommand{\BA}{\langle \mathsf{A}_0\rangle}
\newcommand{\Tr}{\mathrm{Tr}}
\newcommand{\STr}{\mathrm{STr}}
\newcommand{\tr}{\mathrm{tr}}

\newcommand{\I}{\mathrm{i}}
\newcommand{\be}{\begin{eqnarray}}
\newcommand{\ee}{\end{eqnarray}}

\newcommand{\nn}{\nonumber }

\newcommand{\Gk}{\Gamma_k}
\newcommand{\Gt}{\Gamma_k^{(2)}}
\newcommand{\Tc}{T_{\text{c}}}

\newcommand{\Nc}{N_{\text{c}}}
\newcommand{\lqcd}{\Lambda_{\text{QCD}}}
\newcommand{\pat}{\partial_t}
\newcommand{\eq}{\eqref}
\newcommand{\nul}{\nu_\ell}

\begin{document}

\title{On the Nature of the Phase Transition\\ in $SU(N)$, $Sp(2)$ and $E(7)$ Yang-Mills theory}


\author{Jens Braun}
\author{Astrid Eichhorn}
\author{Holger Gies}
\affiliation{Theoretisch-Physikalisches Institut, Friedrich-Schiller-Universit\"at Jena, D-07743 Jena, Germany}
\author{Jan M.~Pawlowski}
\affiliation{Institut f\"ur Theoretische Physik, University of Heidelberg, 
Philosophenweg 16, 62910 Heidelberg, Germany}
\affiliation{ExtreMe Matter Institute EMMI, GSI Helmholtzzentrum f\"ur 
Schwerionenforschung, Planckstr. 1, 64291 Darmstadt, Germany} 

\begin{abstract}
  We study the nature of the confinement phase transition in $d=3+1$
  dimensions in various non-abelian gauge theories with the approach
  put forward in \cite{Braun:2007bx}. We compute an order-parameter
  potential associated with the Polyakov loop from the knowledge of
  full 2-point correlation functions. For $SU(N)$ with $N=3,\dots,12$
  and $Sp(2)$ we find a first-order phase transition in agreement with
  general expectations.  Moreover our study suggests that the phase
  transition in $E(7)$ Yang-Mills theory also is of first order. We
  find that it is weaker than for $SU(N)$. We show that this can be
  understood in terms of the eigenvalue distribution of the order
  parameter potential close to the phase transition.
  \end{abstract}

\maketitle

\section{Introduction}
Understanding the confinement of gluons is a challenging problem,
because characteristic and significant quantities providing both
analytical and numerical access are difficult to identify
straightforwardly.  In recent years, the study of the confinement
mechanism based on the IR behavior of gauge-dependent correlation
functions has turned out to be both fruitful and inspiring. Most
prominently, this has lead to two different confinement scenarios: the
Gribov-Zwanziger \cite{Gribov:1977wm,Zwanziger:1993dh} scenario on 
one hand and the Kugo-Ojima scenario on the other hand
\cite{Kugo:1979gm}. This scenarios are mutually related and
connect signatures of color confinement with the low-momentum behavior
of gluon and ghost correlation functions. They have been
intensely investigated by a variety of nonperturbative field
theoretical tools such as lattice gauge theory \cite{Bonnet:2000}
as well as functional methods
\cite{vonSmekal:1997is,Zwanziger:2001kw,Pawlowski:2003hq,Fischer:2008uz}.

Even if color confinement is eventually properly accounted for by a
corresponding IR behavior of correlation functions, indicating, e.g,
positivity violation and the absence of colored states in the physical state
space, the remaining pressing open question is the relation of color
confinement to quark confinement. 
Typical quark-confinement criteria such as
those based on the Wilson-loop or Polyakov-loop expectation value
\cite{Polyakov:1978vu} in quenched QCD have long remained
inaccessible from the pure knowledge of low-order correlation functions of the
gauge sector. For instance, a direct computation of the heavy-quark potential
requires additional knowledge, e.g., of the quark-gluon
vertex~\cite{Alkofer:2008tt} or of the static gluon correlator in the
Coulomb gauge~\cite{Cucchieri:1996ja}.

In~\cite{Braun:2007bx} a more direct relation between color
  confinement and quark confinement has been established. There, a
  thermodynamic order parameter, namely the Polyakov loop, has been
  computed from the Yang-Mills propagators, thereby establishing for
  the first time a direct link between the IR behavior of correlation
  functions and quark confinement. Aside from the above relation of
  color confinement and quark confinement, the relation between quark
  confinement and chiral symmetry breaking has more recently been
  studied with the aid of so-called dual
  observables~\cite{Gattringer:2006ci,Braun:2009gm}. These
  formal and computational advances better our understanding of the
 mechanisms underlying confinement and chiral symmetry breaking
  and pave the way towards a first principle access to QCD with
  functional methods.  

  In particular, the approach put forward in~\cite{Braun:2007bx} gives
  us access not only to an order parameter for confinement, the
  Polyakov loop, but also to its full effective potential. The latter
  is a crucial input in Polyakov loop extended effective models such
  as the PNJL and the PQM models~\cite{Fukushima:2003fw}. In these
  models, the Polyakov loop potential is an external input the
  parameter of which are fixed to reproduce pure Yang-Mills lattice
  results. Evidently, these physics constraints do not completely fix
  the potential, leave aside its extension to full QCD. Different
  potentials have been studied, for a comparison see e.g.\
  \cite{Schaefer:2009ui}, and the physics at finite chemical potential
  is in fact very sensitive to parameter changes. In this regard the
  present approach provides the opportunity for a qualitative
  improvement of the above models since it allows to fix the
  Yang-Mills potentials completely from first principles, see also
  \cite{Marhauser:2008fz}.  This has motivated further studies such as
  the delevopment of a framework for the study of QCD with two colors
  \cite{Kondo:2010ts}.  Moreover the present approach underlies the 
  fully dynamical continuum study of two-flavor QCD at finite
  temperature and quark chemical potential put forward in \cite{Braun:2009gm}.

The present work is devoted to a more detailed analysis of the
interrelation between color confinement and quark confinement as
deduced in \cite{Braun:2007bx}.  In addition to providing a more
comprehensive technical insight into the underlying ideas, we
illustrate this interrelation and the generality of the approach by
applying it to a variety of non-abelian gauge theories near the
deconfinement phase transition. In fact, while the phase transition is
of second order in $SU(2)$ Yang-Mills theory, it is well-known from
lattice simulations that a first-order phase transition occurs in
$SU(N)$ gauge theories with $N\geq 3$. This brings up the question on
how the nature of the phase transition is related to the properties of
the underlying gauge group. It has been conjectured in
\cite{Svetitsky:1982gs} based on the order-disorder nature of the
deconfinement phase transition with respect to center symmetry that
the phase transition of $SU(2)$ Yang-Mills theory should fall into the
Ising universality class. This observation based on the symmetry
properties of the center of the gauge group does not necessarily
extend to the other gauge groups with a center symmetry agreeing with
a 2nd order universality class. In~\cite{Pepe:2006er,Holland:2003kg}
it has been conjectured that the dynamics near the critical
temperature is sensitive to the mismatch of the number of dynamical
degrees of freedom in the confined and deconfined phase. In accordance
with this conjecture a first-order phase transition has been found for
$Sp(2)$ even though the center of the group is
$Z(2)$~\cite{Pepe:2006er,Holland:2003kg}.  In the present paper we
study $Sp(2)$ and $E(7)$ gauge theory (the center of both groups is
$Z(2)$) and compare the results to our findings for $SU(N)$ gauge
theories also in order to shed more light on the conjecture put
forward in~\cite{Pepe:2006er,Holland:2003kg}.

Our study of the deconfinement phase transition is based on an order parameter
related to the Polyakov loop variable,
\begin{equation}
L[\mathsf{A}_0] =\frac{1}{\Nc} \tr\, \CP\, \exp \left( \I g
  \int\limits_0^\beta dx_0\, \mathsf{A}_0(x_0,x) \right), \label{1}
\end{equation}
more precisely on the expectation value of $\BA$ in Polyakov gauge,
see also \cite{Marhauser:2008fz}.  Then, $\langle{A}_0\rangle$ is
sensitive to topological defects related to confinement
\cite{Reinhardt:1997rm}, and also serves as a deconfinement order
parameter.

The
effective potential of $\BA$ is accessible from the knowledge of gauge
correlation functions by means of the functional renormalization group
(RG), \cite{Braun:2007bx}. As an additional characteristic ingredient for a quantitative
understanding of the phase transition we introduce and identify eigenvalue
distributions  of the order parameter which exhibit characteristic
traces and facilitate a quantitative understanding of the behavior of the
corresponding effective potential. 

The paper is organized as follows: In Sect.~\ref{sec:flows}
and~\ref{sec:backflows} we discuss general aspects of functional flows for a
study of non-abelian gauge theories. In Sect.~\ref{sec:backland} we discuss
how background-field RG flows can be constructed from RG flows in Landau-gauge
Yang-Mills theories.  We discuss a sufficient confinement criterion in
Sect.~\ref{sec:confcrit} before we present our study of the nature of the
phase transition in $SU(N)$, $Sp(2)$ and $E(7)$ gauge theory in
Sect.~\ref{sec:numresults}.

\section{Functional flows and optimization}\label{sec:flows}

For our study of the phase transition of non-abelian gauge theories we employ
the functional RG for the effective action $\Gk$ \cite{Wetterich:1993yh}. This
allows us to interpolate between the initial UV action related to the
classical action $\Gamma_{k=\Lambda}\simeq S$ and the full quantum effective
action $\Gamma\equiv\Gamma_{k=0}$, being the 1PI generating functional. The
infrared (IR) regulator scale $k$ separates the fluctuations with momenta
$p^2\gtrsim k^2$ which are already included in $\Gk$, from those with smaller
momenta which still have to be integrated out. The full RG trajectory is given
by the solution to the Wetterich equation ($t=\ln (k/\Lambda)$),
\begin{equation}
\pat\Gk[\Phi]=\frac{1}{2} \STr \0{1}{\Gt[\Phi]+R_k} \pat R_k\,,
\label{floweq}
\end{equation}
where $\Gt$ denotes the second functional derivative with respect to
the dynamical field $\Phi$, collectively summarizing gluon and ghost
fields in the present context. The super trace $\STr$ sums over
momenta, internal indices and species of fields and includes a
negative sign for the ghost fields.  The regulator function $R_k$
specifies the details of the Wilsonian momentum-shell
integrations. See \cite{Litim:1998nf,Pawlowski:2005xe,Gies:2006wv,Igarashi:2009tj} for
reviews on gauge theories.

In the present context, we are interested in the effective potential for an
order-parameter quantity which is related to the local part of the full
effective action $\Gamma=\Gamma_{k=0}$. The latter can formally be obtained
from the integrated flow,
\begin{equation}
\Gamma[\Phi]=\Gamma_\Lambda[\Phi]-\int_0^\Lambda \0{d k}{k}
\frac{1}{2} \STr \0{1}{\Gt[\Phi]+R_k} \pat R_k\,.
\label{eq:Iflow}
\end{equation}
Eq.~\eqref{eq:Iflow} is an equation for the full quantum effective action
$\Gamma$ and has {\it a priori} no $\Lambda$ dependence.  A
partial integration leads to
\begin{eqnarray}\nonumber 
\Gamma=\frac{1}{2}\STr\ln \Gamma^{(2)}&+&
\int_0^\Lambda\frac{dk}{k} \Delta \Gamma_k
\\ & 
+& \Gamma_\Lambda-
\frac{1}{2}\STr\ln (\Gamma_\Lambda^{(2)}+R_\Lambda)\,. 
\label{eq:iIflowpart}
\end{eqnarray} 
Note that the first term on the right-hand side does not depend on the
regulator function. The initial conditions of the flow at $\Lambda$ including 
possible subtractions are comprised in the second line of Eq.~\eqref{eq:iIflowpart}. 
The second term in the first line of \eqref{eq:iIflowpart} reads 
\begin{eqnarray} 
\Delta \Gamma_k :=-\frac{1}{2}\STr \0{1}{\Gt+R_k} \pat \Gt \,.
\label{eq:improve}
\end{eqnarray} 
For general regulators, $\Delta\Gamma_k$ is only finite upon subtractions
contained in the second line of \Eqref{eq:iIflowpart}, as $\pat \Gt$ does not
vanish for large momenta. Whereas the representation \eqref{eq:iIflowpart}
thus is of limited practical use in the general case, it is ideally suited for
the determination of an order-parameter potential which is UV finite from the
beginning, as is the Weiss potential. In fact, our quantitative results for
the order-parameter potential are dominated already by the
first term of \Eqref{eq:iIflowpart}. For the remainder of this section, we
will, however, be concerned with an optimized strategy to evaluate the
contributions from the $\Delta \Gamma_k$ term.

The evaluation of this `RG-improvement term'
$\sim\partial_t\Gamma_k$ requires two nontrivial ingredients:
full gluon and ghost propagators in the presence of an IR regulator,
$G_k= (\Gamma^{(2)}_k+R_k)^{-1}$ and the flow of the inverse
propagator $\pat \Gt$. This information has been made available in
\cite{jan,Fischer:2008uz} where the full momentum dependence of
Landau-gauge propagators has been computed within optimized RG
flows. For earlier RG calculations, acquiring partial knowledge about
Yang-Mills propagators and providing evidence for the
Kugo-Ojima/Gribov-Zwanziger confinement scenarios,
see~\cite{Ellwanger:1995qf,Pawlowski:2003hq,Fischer:2004uk,Fischer:2008uz}.

Optimization of the RG flow has not only the advantage of a more
stable and faster convergent numerical scheme; in a given truncation,
it can actually be posed as a stability/convergence problem
\cite{Litim:2000ci}. It also provides for a
link to using propagators obtained from lattice simulations, see
below. Full functional optimization can be reformulated as the quest
for a minimal flow trajectory for general functional flows
\cite{Pawlowski:2005xe}. In other words, for a given gap $1/k_{\rm
  eff}^2$ of the propagator which constitutes the inverse of the
physical infrared cut-off, the integrated optimized flow is already as
close as possible to the full theory, as the remaining flow trajectory
is minimal. This results in the following propagator for the
corresponding optimal regulator $R_{\rm opt}$ \cite{Pawlowski:2005xe}:
\begin{eqnarray}\nonumber 
 \0{1}{\Gt+R_{\rm opt}}(p^2) &=&\0{1}{\Gamma_0^{(2)}(p^2)}
\theta(\Gamma_0^{(2)}(p^2)-k_{\rm eff}^2)\\ 
&&+\0{1}{k_{\rm eff}^2}
\theta(k_{\rm eff}^2-\Gamma_0^{(2)}(p^2)).
\label{eq:optprop}\end{eqnarray}
with $k_{\rm eff}^2= \Gamma_0^{(2)}(k^2)$. The propagator
$(\Gt+R_{\rm opt})^{-1 }$ is already the full propagator for all
eigenvalues of $\Gamma_0^{(2)}$ belonging to ${\rm spec}\{
\Gamma_0^{(2)}\}\geq k_{\rm eff}^2$, and is identical to the gap for
the remaining eigenvalues, ${\rm spec}\{\Gamma_0^{(2)}\}<k_{\rm
  eff}^2$.  The choice \eqref{eq:optprop} requires non-trivial field
redefinitions. For the gauge field we have $A=Z^{1/2}_A \hat A$ with
$\partial_t \hat A=0$ and
\begin{eqnarray}\nonumber 
&& \hspace{-1.5cm} \frac{\partial_t Z_A}{Z_A} =  \left(\frac{1}{2} 
\STr \0{1}{\Gt+\!R_k} \pat R_k\!\right)^{(2)} (p^2)\\[1ex]
& & \hspace{1.5cm}\times \0{1}{\Gamma_0^{(2)}(p^2)} 
\theta(k_{\rm eff}^2\!-\!\Gamma_0^{(2)}(p^2))\,,\label{eq:wavefunc}
\end{eqnarray}
where the right-hand side is evaluated at $\Phi=0$. 
Eq.~\eqref{eq:wavefunc} ensures $\partial_t \Gamma_k^{(2)} \theta(k_{\rm
  eff}^2-\Gamma_0^{(2)})=0$, and hence
\begin{equation}
( \Gamma_k^{(2)}-
\Gamma_0^{(2)})\theta(k_{\rm eff}^2-\Gamma_0^{(2)})=0\,.
\end{equation}
The conditions~\eqref{eq:optprop} and \eqref{eq:wavefunc} allow us to provide the
optimal regulator in an explicit form \cite{Pawlowski:2005xe}:
\begin{eqnarray}
R_{\rm opt}=(k_{\rm eff}^2-
\Gamma_k^{(2)}(p^2))\theta(k_{\rm eff}^2-\Gamma_0^{(2)}(p^2)).
\label{eq:optreg}
\end{eqnarray}
With the choice \eqref{eq:optreg}, the flow of Green's functions can be
computed within an iteration of the integrated flow starting from an initial
value for $\Gamma_0^{(2)}$. 

Let us elucidate aspects of the optimized flow in the context of the
integrated flow \eqref{eq:iIflowpart}. For the optimized flow the
relation \eqref{eq:iIflowpart} follows from a direct integration of
the flow: the first term in \Eqref{eq:iIflowpart} relates to
integrating the $\partial_t k^2_{\rm eff}$ contributions of the
related flow, the second term in \Eqref{eq:iIflowpart} is the $t$
integral of the contributions $\sim\partial_t\Gamma_k^{(2)}$. Details
of the numerical computation of Eq.~\eqref{eq:iIflowpart} can be found
in App.~\ref{App:Details}.

From a general perspective, the effective action~\eqref{eq:iIflowpart}
together with the optimized regulator can be considered as a DSE within a
consistent BPHZ-type non-perturbative renormalization~\cite{Pawlowski:2005xe},
where the $\Lambda$-dependent terms provide the classical action and the
subtraction terms. The computational benefit in comparison to standard DSE
equations is the explicit finiteness of \Eqref{eq:iIflowpart} in any
truncation without the need of further additive or multiplicative
renormalizations. The second term on the right-hand side of the first line  
constitutes an RG improvement term.

\section{Background field flows}\label{sec:backflows}

In order to arrive at the effective potential for an order-parameter field, we
parameterize the fluctuations with respect to a background field which is
related to the order parameter. In Yang-Mills theories, this decomposition
into fluctuating modes and the background field can be organized such that the
resulting background-field action preserves a residual gauge symmetry, e.~g.
\cite{Abbott:1981hw}.  This approach using the background-field gauge can be
understood as a simple extension of Yang-Mills theories within general
covariant gauges. The gauge condition $\partial_\mu A_\mu=0$ is generalized to
\begin{equation}\label{eq:backgauge}
D_\mu(\bar A) (A-\bar A)_\mu=0,
\end{equation}
for an unspecified background field $\bar A$. Equation \eqref{eq:backgauge}
implemented on configuration space in a strict sense defines Landau-DeWitt
gauge. A less strict Gau\ss ian average over the gauge condition $D_\mu(\bar
A) (A-\bar A)_\mu=\CC$ with a probability distribution $\sim \exp ( -1/\xi \int
\tr\, \CC^2)$ leads to the background-field equivalent of a general covariant
gauge.  Such a formulation has the benefit of an auxiliary gauge symmetry for
the effective action under a transformation of both, the full gauge field
$A\to A+D\omega$ and the background $\bar A\to \bar A+\bar D\omega$. In this
manner, the gauge condition \eqref{eq:backgauge} is unchanged since $a=A-\bar A$
transforms as a tensor, $a\to [a,\omega]$.  

With the gauge fixing \eqref{eq:backgauge}, the effective action now depends
also on the auxiliary field, $\Gamma=\Gamma[\Phi,\bar A]$ with $\Phi=(a,C,\bar
C)$. We emphasize that the background-field gauge transformation is an
auxiliary symmetry. The effective action $\Gamma=\Gamma[\Phi,\bar A]$ still
carries non-trivial symmetry constraints, namely the Slavnov-Taylor identities
(STI). These follow from a gauge or BRST transformation of the field $\Phi$ at
fixed background. Indeed, the underlying STI are that of a standard covariant
gauge. Even though the gauge invariance is an auxiliary symmetry, it
facilitates the construction of a (physically) gauge-invariant effective
action $\Gamma[A]=\Gamma[0,\bar A= A]$.  The flow equation for
$\Gamma[\Phi,\bar A]$ in such a setting reads
\begin{equation}
\pat\Gk[\Phi,\bar A]=\frac{1}{2} \Tr \0{1}{\Gamma_k^{(2,0)}
[\Phi,\bar A]+R_k} \pat R_k\,,
\label{eq:backflow}
\end{equation}
where 
\begin{eqnarray}\label{eq:nm}
\Gamma_k^{(n,m)}=\0{\delta^{n} }{\delta\Phi^n}\0{\delta^{m}}{
\delta\bar A^m}\Gamma_k\,.
\end{eqnarray}  
The action $\Gk[\Phi,\bar A]$ is still gauge invariant under background gauge
transformations provided the regulator transforms as a tensor under gauge
transformations,
\begin{eqnarray}\label{eq:Rinv} 
R_k\to [R_k,\omega]\,.
\end{eqnarray} 
This can be established by using background-covariant momenta $p\to
-i D(\bar{A})$ in the regulators, or $p^2\to \Gamma^{(2)}[0,\bar A]$ as necessary
for optimized flows. For $\Phi=0$, \Eqref{eq:backflow} entails the flow of
$\Gamma_k[A]$,
\begin{equation}
\pat\Gk[A]=\frac{1}{2} \STr \0{1}{\Gamma_k^{(2,0)}
[0,A]+R_k} \pat R_k\,,
\label{eq:backflow0}
\end{equation}
which is gauge invariant for regulators obeying \Eqref{eq:Rinv}.  Note
that the flow \eqref{eq:backflow0} is not closed
\cite{Pawlowski:2005xe,Pawlowski:2001df,%
  Litim:2002ce}: the right-hand side depends on
$\Gamma_k^{(2,0)}[0,A]$, whereas the left-hand side only allows for
the computation of $\Gamma_k^{(2)}=\Gamma_k^{(0,2)}[0,A]$. The
necessary approximation for the direct use of \eqref{eq:backflow0} is
therefore to set
\begin{eqnarray}\label{eq:nectrunc} 
\Gamma_k^{(2,0)}[0,A]\stackrel{!}{=}\Gamma_k^{(0,2)}[0,A]\,.
\end{eqnarray}
The resulting flow \eqref{eq:backflow0} is indeed closed and can be
solved within powerful heat-kernel techniques
\cite{Reuter:1993kw,Gies:2002af,Litim:2002ce,Braun:2005cn,Manrique:2010am}.
For instance, this approach predicts the existence of an infrared
fixed point of the coupling at zero and finite temperature
\cite{Gies:2002af}, as it is similarly found in Landau gauge QCD.

However, within the present context of the order-parameter potential, it is
crucial to go beyond the approximation \eqref{eq:nectrunc} used in 
\cite{Braun:2005cn}, as the confinement-deconfinement phase transition is
rather sensitive to the correct mid-momentum and infrared behavior of the
fluctuation-field propagator $\Gamma_k^{(2,0)}[0,A]$.  Therefore, a proper
distinction between the fluctuation-field and background-field dependence of
the action is mandatory.

\section{Background-field flows and Landau-gauge Yang-Mills
  theory}\label{sec:backland}

The crucial ingredient for our studies of the confinement phase
transition is the two-point function $\Gamma_k^{(2,0)}[0,A]$. As it is
not the output of the pure background-field flow \eqref{eq:backflow0},
we have to compute it separately. From now on, we restrict ourselves
to the Landau-DeWitt gauge~\eqref{eq:backgauge} with the gauge
parameter set to $\xi=0$. This gauge has several benefits: first, it
projects on (covariantly) transversal degrees of freedom, and second,
the longitudinal components of Green functions decouple from the
dynamics of the transversal ones, and thirdly it is a fixed point of
the flow \cite{Litim:1998qi}. As the longitudinal components are subject to
modified Slavnov-Taylor identities, this minimizes the truncation
error. This is also related to a second issue, namely the gauge
dependence of the background-field effective action $\Gamma_k[A]$.
For Landau-DeWitt gauge, this action is identical to the geometrical
effective action, see e.g.
\cite{Branchina:2003ek,Pawlowski:2003sk,Pawlowski:2005xe}, which is
gauge invariant also with respect to quantum gauge
transformations. There, the fluctuation field agrees with $a=A-\bar A$
only in leading order. The background-field approach in Landau-DeWitt
gauge can indeed be understood as the leading order of a manifestly
gauge-invariant approach to functional RG flows
\cite{Pawlowski:2003sk,Pawlowski:2005xe}.

We proceed with constructing the key input $\Gamma_k^{(2,0)}[0,A]$. 
First we remark that $\Gamma_k^{(2,0)}[0,0](p^2)$ is simply the
propagator in Landau gauge which has been computed on the lattice
\cite{Bonnet:2000} as well as by functional methods
\cite{vonSmekal:1997is,Zwanziger:2001kw,Pawlowski:2003hq,Fischer:2008uz}. 
The full RG trajectory $\Gamma_k^{(2,0)}[0,0](p^2)$ 
has been computed in \cite{jan,Fischer:2008uz} for the optimized regulator $R_{\rm opt}$.
Now the auxiliary background gauge symmetry comes to our aid. It
constrains the extension of the Landau-gauge two-point function 
$\Gamma_k^{(2,0)}[0,0]$ to $\Gamma_k^{(2,0)}[0,A]$ as the latter has 
to transform as a tensor under gauge transformations. We conclude that 
\begin{equation}\label{eq:genprop}
(\Gamma_{k}^{(2,0)}[0,A])^{ab}_{\mu\nu}=(\Gamma_{k}^{(2,0)}[0,0]
(-D^2) )^{ab}_{\mu\nu}+
F^{cd}_{\rho\sigma} f_{\mu\nu\rho\sigma}^{abcd}(D)\,,
\end{equation}
where $F^{cd}_{\rho\sigma}$ denotes the field strength tensor in the adjoined
algebra, and the function $f(x)$ is non-singular at $x=0$. Note that covariantly
longitudinal correction terms in \Eqref{eq:genprop} are irrelevant as we are
in the Landau-DeWitt gauge. The $f$ terms cannot be obtained from the Landau gauge
propagator. They are indeed related to higher Landau-gauge Green functions.

Next we briefly recall the results for the Landau gauge
propagators~\cite{Pawlowski:2003hq,Fischer:2004uk,Fischer:2008uz,jan}:
the ghost and gluon propagators can be parameterized as
\begin{eqnarray}\label{eq:gluon} 
\Gamma_{k,A}^{(2,0)}[0,0](p^2)&=&p^2 Z_A(p^2) \Pi_{\rm T} (p) \id \nonumber\\
&& \qquad +\, p^2 \0{Z_{\text{L}}(p^2)}{\xi} \Pi_{\rm L}(p) \id \,,
\end{eqnarray} 
where 
\begin{eqnarray}\label{eq:projections} 
&&\left(\Pi_{\rm T}\right)_{\mu\nu}(p)=\delta_{\mu\nu}-\0{p_\mu p_\nu}{p^2}\,,\;
\left(\Pi_{\rm L}\right)_{\mu\nu}(p)=\0{p_\mu p_\nu}{p^2}\,,\nonumber\\
&&\qquad\qquad\qquad\qquad\quad\id_{ab}=\delta_{ab}\,, 
\end{eqnarray} 
for the gluon and 
\begin{eqnarray}\label{eq:ghost} 
\Gamma_{k,C}^{(2,0)}[0,0](p^2)=p^2 Z_C(p^2)\id\,
\end{eqnarray} 
for the ghost. For the longitudinal dressing function, we have
$Z_{\text{L}}=1+\CO(\xi)$.  Hence, it drops out of all diagrams beyond one
loop. In the deep infrared, the dressing functions $Z_{A,C}$ exhibit a leading
momentum behavior
\begin{eqnarray}\label{eq:ir}
Z_A(p^2\to 0)\simeq (p^2)^{\kappa_A}\,,\quad
Z_C(p^2\to 0)\simeq (p^2)^{\kappa_C}.
\end{eqnarray} 
Landau-gauge Yang-Mills theory admits a one-parameter family of infrared
solutions consistent with RG invariance \cite{Fischer:2008uz}, the underlying
structure still being subject to current research. This family of solutions
can be parameterized by an infrared boundary condition for the ghost
propagator, specifying a value for $Z_C(p^2=0)$.  This fact is reflected in
recent lattice solutions on relatively small lattices, \cite{Maas:2009se}, 
and in the strong-coupling limit~\cite{Sternbeck:2008mv}, for an alternative point of view see 
\cite{Cucchieri:2009zt}. 

\begin{figure*}[t]
  \hspace*{0.0cm}
\includegraphics[width=0.41\linewidth]{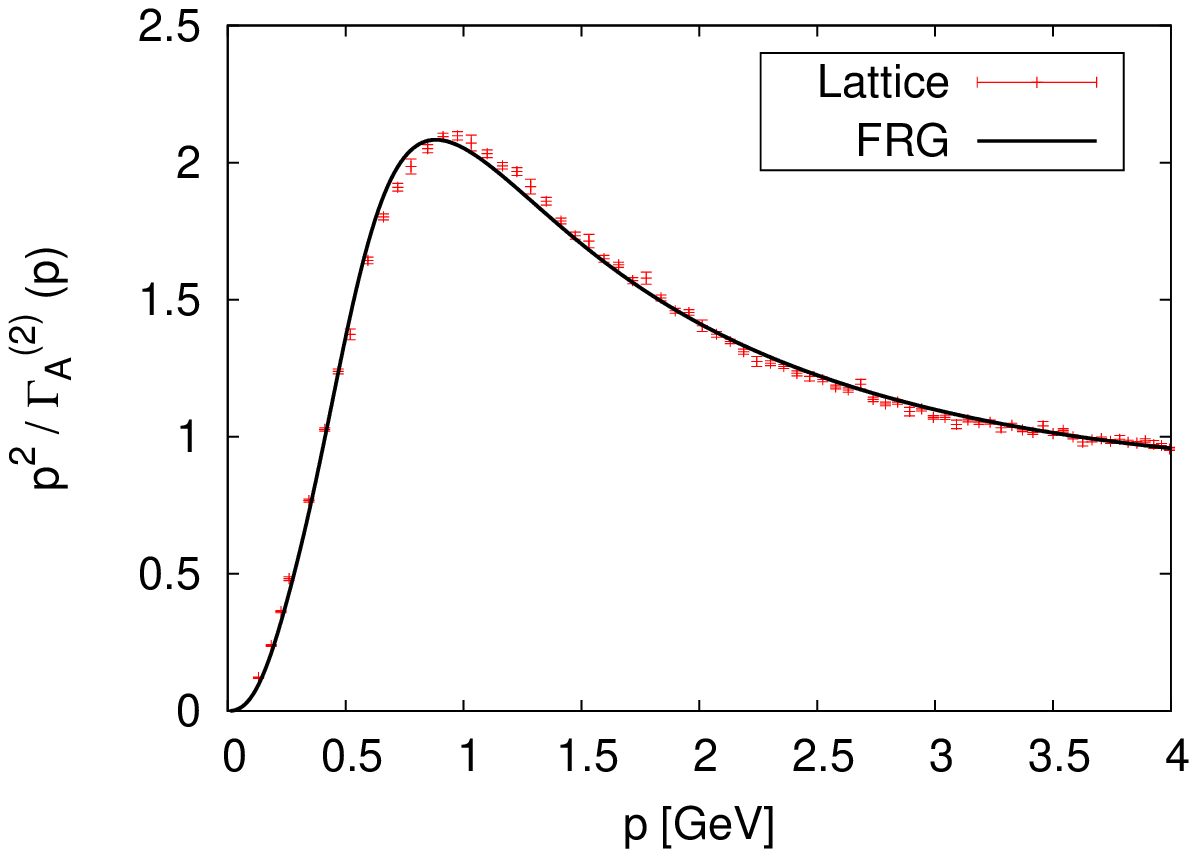}
  \hspace*{2.5cm}
\includegraphics[width=0.41\linewidth]{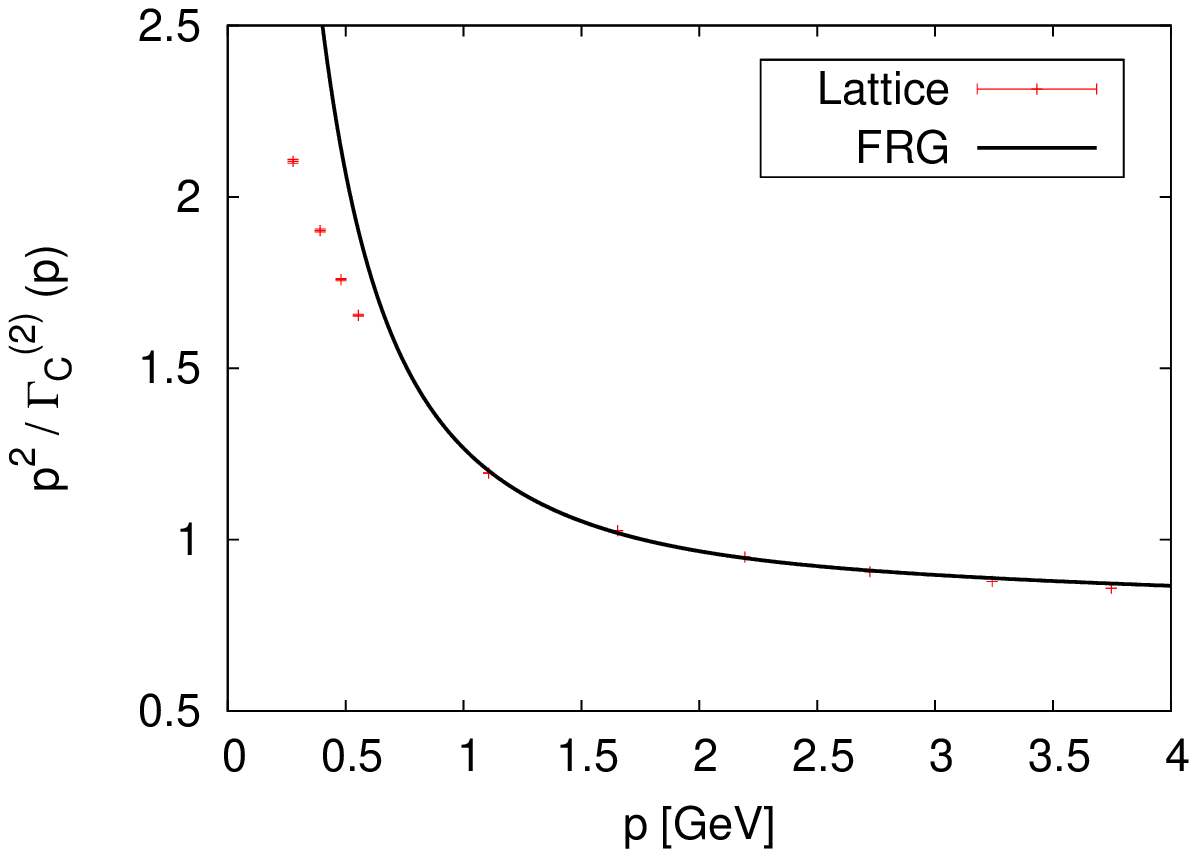} 
\caption{Momentum dependence of the gluon (left panel) and ghost (right panel) 
2-point functions at vanishing temperature. We show the FRG results from 
Ref.~\cite{Fischer:2008uz} (black solid line) and from lattice simulations from 
Ref.~\cite{Bonnet:2000} (red points).}
\label{fig:propagators}
\end{figure*}

For $Z_C(p^2 \to 0)\to 0$, there is a unique {\em scaling} solution,
\cite{Fischer:2006vf,Alkofer:2008jy}. The two exponents $\kappa_A$ and
$\kappa_C$ are then related by the sum rule arising from a non-renormalization
theorem for the ghost-gluon vertex \cite{Taylor:1971ff},
\begin{equation}
  0=\kappa_A + 2\kappa_C +\frac{4-d}{2},\label{eq:sumrule}
\end{equation}
in $d$ dimensional spacetime
\cite{Lerche:2002ep,Zwanziger:2001kw,Fischer:2006vf}. Admissible solutions are
bound to lie in the range $\kappa_C\in [1/2\,,\,1]$. In pure non-abelian gauge
theories, $\kappa_C$ has been computed by a variety of
methods~\cite{Bonnet:2000,vonSmekal:1997is,Zwanziger:2001kw,Pawlowski:2003hq,Fischer:2008uz}.
The precise value depends on the IR behavior of the ghost-gluon
vertex~\cite{Lerche:2002ep}. In most DSE and FRG computations we are led to ($d=4$)
\begin{eqnarray}\label{eq:scaling}  
\kappa_C=0.59535...\quad {\rm and} \quad \kappa_A=-2\kappa_C=- 1.1907..., 
\end{eqnarray} 
being the value for the optimized regulator \cite{Pawlowski:2003hq}.  The
regulator dependence in FRG computations leads to a $\kappa_C$ range of
$\kappa_C\in [0.539\,,\, 0.595]$, see \cite{Pawlowski:2003hq}; for a
specific flow, see \cite{Fischer:2004uk}. These results entail the
Kugo-Ojima/Gribov-Zwanziger confinement scenario. The gluon is infrared
screened, the propagator even tends to zero, see
\eqref{eq:ir},\eqref{eq:scaling}, whereas the ghost is infrared enhanced. Due to
the non-renormalization property of the ghost-gluon vertex,
\cite{Lerche:2002ep,Taylor:1971ff}, a running coupling can be defined in terms
of
\begin{eqnarray}\label{eq:alphas} 
\alpha_s(p^2) = \0{g^2}{4 \pi Z_A(p^2) Z_C^2(p^2)}\,, 
\end{eqnarray}
which runs towards an IR fixed point, see Eq.~\eqref{eq:ir}. In
Fig.~\ref{fig:propagators}, we show the momentum dependence of the ghost- and
gluon propagator as obtained from a functional RG study~\cite{Fischer:2008uz}
in comparison to lattice results \cite{Bonnet:2000}.

A different type of {\em decoupling} solution is found for non-vanishing
$Z_C(0)$: here, the gluon propagator tends to a constant in the infrared, $
p^2 Z_A(p^2) \to m^2$, for related work see e.g.\
\cite{Fischer:2008uz,Cornwall:1981zr}. It should be stressed that the
gluon propagator then does not correspond to the propagator of a massive
physical particle, but clear indications for positivity violation related to
gluon confinement are observed, \cite{Fischer:2008uz,Cucchieri:2004mf}. Still,
the gluon decouples from the dynamics as does a massive particle. The
qualitative infrared behavior is then characterized by the exponents
\begin{equation}
  \kappa_A=-1\,, \quad {\rm and} \quad 
  \kappa_C=0\,.\label{eq:decoupling}
\end{equation}
Though the infrared exponents for the scaling \eq{eq:scaling} and decoupling
solutions \eq{eq:decoupling} deviate from another, the propagators do only
differ in the deep infrared.  It has been suggested in \cite{Fischer:2008uz}
that the infrared boundary condition is directly related to the global part of
the gauge fixing, hence reflecting different resolutions of the Gribov
problem.  Indeed, the infrared boundary condition has been used as a global
completion of the gauge fixing in \cite{Maas:2009se}.

For most parts of the present work, the difference between the scaling and the
decoupling solution is of minor importance. For concrete numerical
computations, we combine information about the propagators as obtained from
the lattice as well as functional methods, completing these propagators in the
deep IR with the scaling solution. The latter is actually singled out by the
requirement of global BRST for Landau gauge Yang-Mills with standard local
BRST invariance.

The discussion so far applies to Yang-Mills theory at zero
temperature. Several modifications arise in the presence of a thermal bath,
such as the immediate replacement of continuous loop energies $p_0$ by
Matsubara frequencies $p_0\to\omega_n=2\pi nT$ in the imaginary-time
formalism. Moreover, the gluon propagator acquires an additional component, as
the contributions longitudinal and transversal to the heat bath become
independent. 

In this work, we neglect the finite-temperature modifications of the
propagators, but work with zero-temperature propagators evaluated at the
Matsubara frequencies. In scalar theories it has been shown that this
approximation already provides a quantitative insight into the
finite-temperature phase structure~\cite{Braun:2009si}.  First results for
finite-temperature gluon- and ghost-propagators indeed indicate that the
propagators are little
modified~\cite{Gruter:2004bb,Fischer:2010fx} for 
Matsubara frequencies $2 \pi T n$ with $|n|\gtrsim 2,3$. Significant changes have
been found for the gluon propagator longitudinal to the heat bath which is
increased compared to the transversal counterpart.  We stress that the
inclusion of the full temperature dependence of the propagators as well as the
order-parameter fluctuations is inevitable for an accurate determination of,
e.g., the critical exponents or the thermodynamic properties of the theory
(see, e.g., \cite{Chernodub:2007rn}).

The presence of finite temperature also takes influence on the form of the
propagator at finite background field, as another field invariant, the
Polyakov loop $L$, exists. This adds further terms to the right-hand side of
\Eqref{eq:genprop}. Specializing to constant background fields $\bar{A}=\BA$, 
we find schematically
\begin{equation}\label{eq:genpropA0}
(\Gamma_{k}^{(2,0)}[0,\BA])^{ab}_{\mu\nu}=(\Gamma_{k}^{(2,0)}[0,0]
(-D^2) )^{ab}_{\mu\nu}+\,
L\,\,\text{terms}
\,, 
\end{equation} 
since the $f$ term in \eq{eq:genprop} vanishes, $F(\BA)=0$, for
$\BA=$const. In this work we drop the $L$ contribution in
\eq{eq:genpropA0} which is related to the second derivative of the
order-parameter potential via Nielsen
identities~\cite{Pawlowski:2005xe,Pawlowski:2001df}.  We expect that
this term affects our results only on a quantitative level, e.~g., the
quality of critical exponents. For example in $SU(2)$ gauge theory, we
expect a second-order phase transition. Here, the critical dynamics
encoded in the critical exponents is sensitive to order-parameter
fluctuations as is well known from studies of scalar $O(N)$
theories. The $L$ terms take a direct influence on the spectrum of
order-parameter fluctuations, so that we expect these terms to be
relevant at criticality. Indeed, the role of order-parameter
fluctuations has been studied in Ref.~\cite{Marhauser:2008fz} for
$SU(2)$ where the correct $Z_2$ critical exponents have been found.
 However, the phase transition in $SU(N)$
gauge theories ($N\geq 3$) is of first order and therefore less
affected by our approximation of dropping the $L$ terms.

Let us finally stress that our approximations at finite temperature do not
take any influence on our conclusions about confinement in the
zero-temperature limit, discussed below. In particular, the background
covariantization of the transverse propagator in \Eqref{eq:genprop} becomes
exact in this limit, representing a first important result of our present
work. This paves the way for a fully consistent low-energy RG analysis of QCD
in the background-field formalism, see Refs.~\cite{Gies:2002hq,Braun:2008pi}
for a study of 1-flavor QCD. 

From Eq.~\eqref{eq:genprop} and the above
results for the Landau gauge propagators we already conclude that the
truncation \eq{eq:nectrunc} is not working well in the (deep) infrared ($p\ll
\Lambda_{\rm QCD}$).  There we expect a fixed point for the coupling,
\eq{eq:alphas}, which entails a constant dressing for the propagator of the
background field, $\Gamma^{(0,2)}$: background gauge invariance leads to RG
invariance of $ g \bar A$, and hence to
\begin{eqnarray}\label{eq:gaugerel} 
Z_{\bar A}\sim Z_g^{-1}\,, 
\end{eqnarray} 
which results in a constant dressing. Therefore the background field
propagator $1/\Gamma_0^{(0,2)}$ diverges in the infrared whereas the
propagator of the dynamical fluctuation field $a$, $1/\Gamma_0^{(2,0)}$
is suppressed in the infrared for both scaling and decoupling
solution. Moreover, the ghost propagator is infrared enhanced for the
  scaling solution in contradistinction to the one-loop truncation used so
far in most background field flows, that is $Z_C=1$ which would  only be
  compatible with the decoupling solution. We conclude that
important aspects of the infrared physics can easily be missed by truncations
based on \eq{eq:nectrunc}.

\section{Confinement criterion}\label{sec:confcrit}

Our study of the deconfinement phase transition 
is based on an order parameter related to the Polyakov loop variable, see Eq.~\eqref{1}. 
The negative logarithm of the Polyakov loop expectation value
$\langle L \rangle$ can be interpreted as the free energy of a single static
color source in the fundamental representation of the gauge group
\cite{Svetitsky:1985ye}. In this sense, an infinite free energy associated
with confinement is indicated as $\langle L\rangle \to 0$, whereas $\langle
L\rangle\neq 0$ signals deconfinement.  For gauge groups with a nontrivial
center, $\langle L\rangle$ measures whether center symmetry is realized by the
ensemble under consideration \cite{Svetitsky:1985ye}. As $\langle L\rangle$
transforms nontrivially under center transformations a center-symmetric
(disordered) ground state automatically ensures $\langle L\rangle=0$, whereas
deconfinement $\langle L\rangle\neq 0$ is related to the breaking of this
symmetry, pointing to an ordered phase.

The background-field formalism used in this work allows us to fix the
fluctuations with respect to Landau-DeWitt gauge and simultaneously maintain
gauge invariance for the background field $\bar A$ which we relate to the
Polyakov loop in the following manner: we use the Polyakov
gauge~\cite{Reinhardt:1997rm} by gauge-rotating the background field into the
Cartan subalgebra and imposing $\partial_0 \bar{A}_0=0$. From the knowledge of
$\bar{A}_0$, the value of the corresponding Polyakov loop $L[\bar{A}_0]$ can
immediately be inferred. Once the effective action $\Gamma[A=\bar{A}]$ is
constructed within the background-field formalism, the minimum of the action
represents the expectation value of the fluctuating quantum field, $A=\bar{A}=
\langle \mathsf{A} \rangle$. The corresponding Polyakov-loop value then is
$L[A_0] = L[\BA]$. 

In addition to $\langle L\rangle$, also $L[\BA]$ serves as an order parameter
for confinement: first, $L[\BA]$ is an upper bound for the Polyakov loop
expectation value due to the Jensen inequality, $L[\BA]\geq \langle L
[\mathsf{A}_0]\rangle$, and therefore is nonzero in the deconfined
phase. Second, it has been shown in \cite{Marhauser:2008fz} that $L[\BA]$
vanishes identically in the center-symmetric phase where also $\langle L[A_0]
\rangle=0$. We conclude that together with $L[\BA]$ also $A_0=\bar{A}_0=\BA$
is an order parameter for center symmetry and confinement in the Polyakov
gauge. In the following, we indeed concentrate on the effective potential
$V(\BA)$ for this order parameter.

Let us now recapitulate the confinement criterion put forward in
Ref.~\cite{Braun:2007bx}. This criterion relates the IR behavior
of gluon and ghost 2-point functions to the effective potential for the order
parameter $\BA$, starting from the full flow as displayed in
\Eqref{eq:iIflowpart}.

The following simplified analytical discussion is based on the assumption that
the second term in \Eqref{eq:iIflowpart} proportional to $\sim \partial_t
\Gamma_k^{(2)}$ (cf. \Eqref{eq:improve}) is subleading. This term with the
explicit $k$ integral resembles the full integrated flow except for the
substitution $\partial_t R_k\to \partial_t \Gamma_k^{(2)}$. In the UV, its
subleading role is obvious, since $\partial_t \Gamma_k^{(2)}$ is of order
$\alpha_s$ whereas $\partial_t R_k$ is of order one. In the deep infrared such
an ordering cannot be found. Nevertheless, our full numerical study shows that
the term depending on $\partial_t \Gamma_k^{(2)}$ is subleading for a study of
the Polyakov loop on all scales studied in this work.

Anticipating the subdominance of the $\partial_t \Gamma_k^{(2)}$ term, we
study the influence of the first term of Eq.~\eq{eq:iIflowpart} on the
Polyakov-loop potential in the UV and IR regime. (The remaining terms are
irrelevant for this discussion, as the potential is finite and does not
require counterterms.) In the UV regime ($p^2\gg T^2$), perturbation theory
holds and the inverse propagators of the longitudinal and transversal gluons
and the ghosts are given by
$\Gamma^{(2),\text{pert}}_{\text{L,A,gh}}(p^2)=p^2$. In the presence of a
constant background field $\BA$, the momentum is replaced by the background
covariant derivative, i. e. $p_0 \to -\I D_0$. With the parameterization
\begin{equation}
\beta g \langle \mathsf{A}_0^a\rangle= 2\pi\!\!\!\!\!\!\! \sum_{T^a \in
  \rm{Cartan}} \!\!\!\!\!\!T^a\phi^a
= 2\pi\!\!\!\!\!\!\! \sum_{T^a \in \rm{Cartan}}\!\!\!\!\!\! T^a v^a |\phi|,\,\, v^2=1, \label{eq:HermColM}
\end{equation}
the spectrum of the background covariant Laplacian becomes
\begin{equation}
p^2 \to \text{spec}\{-D^2[\BA]\} =\vec{p}^{\,2} + (2\pi T)^2 (n - |\phi|
\nul)^2\,,\label{spec}
\end{equation}
where $n\in \mathbbm{Z}$, and $(T^a)^{bc}=-\I f^{abc}$ denotes the generators
of the adjoint representation of the gauge group under consideration. $\nul$
denotes the eigenvalues of the hermitian color matrix occurring in
\Eqref{eq:HermColM},
\begin{equation}
\nul= \text{spec}\{ (T^a v^a)^{bc} | v^2=1 \}, \label{eq:nul}
\end{equation}
and therefore depends on the direction of the unit vector $v^a$. The index
$\ell$ labels these eigenvalues, the number of which is equal to the dimension
$d_{\rm adj}$ of the adjoint representation of the gauge group, $\ell=1,
\dots, d_{\rm adj}$, e.g., $d_{\rm adj} = N^2-1$
for $SU(N)$. For each non-vanishing eigenvalue $\nul$ there exists an
eigenvalue $-\nul$. For $SU(2)$, we have $\nul=\pm 1,0$. Equation \eqref{spec}
reveals that $\phi^a= |\phi| v^a$ denotes a set of compact variables, as an
arbitrary shift of $\phi^a$ can be mapped back onto a compact domain for
$\phi^a$ by a corresponding shift of $n$, i.e., the Matsubara frequency.

With these prerequisites, the perturbative limit of the effective
order-parameter potential $V$ in $d>2$ dimensions is given by
\begin{eqnarray}
\frac{V^{\text{UV}}(\phi)}{T^d}
\!=\!\frac{(2\!-\!d)\Gamma(\frac{d}{2})}{\pi^{d/2}}\, 
\sum_{l=1}^{d_{\rm{adj}}} \sum_{n=1}^\infty \frac{\cos 2\pi n |\phi| \nul
}{n^d}\!.\label{eq:Weiss}
\end{eqnarray}
Here, we have dropped a temperature- and field-independent constant. The
dimensionality of the potential is determined by the dimension of the Cartan
(sub)algebra. This perturbative $V^{\text{UV}}$ corresponds to the well-known
Weiss potential \cite{Weiss:1980rj}, generalized to $d$
dimensions \cite{Actor:1982uc}. It exhibits maxima at the
center-symmetric points where $L[\BA]=0$ (and thus also $\langle L
\rangle=0$), implying that the perturbative ground state is not confining,
i.e. $\langle L \rangle \neq0$.  Since the eigenvalues $\nul$ are pairwise
identical with respect to their absolute values, the Weiss potential for a
given gauge group can be considered as a superposition of $SU(2)$ potentials
with different periodicities determined by the eigenvalues $\nul$. The
eigenvalues can be viewed as Fourier frequencies of the order-parameter
potential. We stress that this also holds in non-perturbative studies of the
Weiss potential. Hence we have
\begin{equation}
V(\phi)=\frac{1}{2}\sum_{l} V_{\rm SU(2)}(\nu_l|\phi|)\,.\label{eq:Vsum}
\end{equation}

Next we perform the same analysis in the IR. With the
parameterizations~\eq{eq:gluon} and \eq{eq:ghost}, the dressing functions
$Z_A(p^2),Z_C(p^2)$ are characterized by the power-law behavior \eq{eq:ir} in
the deep IR, $p^2\ll \lqcd^2$. Quantitatively, the effective potential
$V(\phi)$ is dominantly induced by fluctuations with momenta near the
temperature scale $p^2\sim (2\pi T)^2$.  At low temperatures $(2\pi T)\ll
\lqcd$, the first term of Eq.~\eqref{eq:iIflowpart} thus induces an effective
potential which arises dominantly from fluctuations in the deep IR,
characterized by the exponents $\kappa_{A,C}$. By coupling the fluctuations to the
background field, $p^2\to -D^2[\BA]$, we obtain the following low-temperature
effective potential from the power-law behavior of the two-point Green
functions in the deep IR:
\begin{eqnarray}
V^{\text{IR}}(\phi)
&=& \left\{ 1+ \frac{(d-1)\kappa_A-2\kappa_C}{d-2} \right\}
V^{\text{UV}}(\phi).\nonumber \label{eq:WeissIR}
\end{eqnarray}
Compared to the perturbative Weiss potential~\eqref{eq:Weiss}
we observe that the effective potential is reversed if
\begin{equation}
2\kappa_C - (d-1) \kappa_A > d-2. \label{eq:confcrit}
\end{equation}
In this case, the confining center-symmetric points of the Weiss potential
turn from maxima to minima: the order parameter acquires a center-symmetric
value, such that $L[\BA]=\langle L \rangle =0$. We conclude that
\Eqref{eq:confcrit} serves as a criterion for quark confinement. Provided that
the term $\sim \Gamma_k^{(2)}$ that we dropped for this discussion does
not modify this result, this criterion is sufficient for the occurrence of a
center-symmetric confining phase at low temperatures. 

Let us discuss this criterion in the light of the IR solutions for the
propagators available in the Landau gauge. For the {\em scaling} solution, the
IR exponents are related by the sum rule \eqref{eq:sumrule}, simplifying the
confinement criterion to
\begin{equation}
\kappa\equiv \kappa_C > \frac{d-3}{4}.
\label{eq:crit}
\end{equation}
It is instructive to compare this simple criterion for quark confinement with
related criteria in the $d=4$ case: $\kappa_{d=4}>1/4$.  This criterion
includes the Kugo-Ojima criterion for color confinement $\kappa>0$ as well as
the Zwanziger horizon condition for the ghost $\kappa>0$, which are both
necessary but not sufficient criteria. The corresponding horizon condition for
the gluon, $\kappa>1/2$, is stronger, since the latter is a sufficient but not
a necessary condition for the transversal gluons to exhibit positivity
violation.

For the {\em decoupling} solution with $\kappa_C=0$ and $\kappa_A=-1$ in $d=4$, the
criterion \eqref{eq:confcrit} is satisfied as well and hence, the whole
one-parameter family of Landau-gauge IR solutions is confining. 

To summarize: in color-confined gauge theories, the suppressed gluon and the
enhanced (or constant) ghost fluctuations in the IR induce an
effective potential for the Polyakov loop which corresponds to a
center-disordered confining ground state; this implies an infinite free energy
for a single quark and thus relates color confinement to quark confinement.

\section{Numerical results}\label{sec:numresults}

In the following, we present our results for the order parameter $L[\BA]$ as a
function of temperature for $SU(N)$, $Sp(2)$ and $E(7)$ Yang-Mills theory.
Recall that $L[\BA]\geq \langle L[\mathsf{A}_0] \rangle$ in the deconfined
phase.  Our computation of the effective potential $V(\BA)$ for $SU(N)$,
$Sp(2)$ and $E(7)$ Yang-Mills theory involves several approximations which can
easily be improved on, once more precise propagator data from the lattice or
from functional methods is available: first, we employ the same solution for
the ghost and gluon propagators as obtained from a functional RG
study~\cite{Fischer:2008uz,jan} for all gauge groups, see also Fig.~\ref{fig:propagators}.
In a first approximation, this can be
justified, since the propagators are identical in leading order in a $1/N$
expansion where $N$ is the number of colors. However, even for a small 
number of colors, it has indeed been found on the lattice that $SU(2)$ and $SU(3)$ 
propagators agree within errors~\cite{Cucchieri:2007zm}. 

As a second approximation, we do not take a possible modification of the
functional form of zero-temperature and finite-temperature propagators into
account. We also neglect that the transversal gluon propagator splits into
independent components longitudinal and transversal to the heat bath at finite
$T$. Our approximation to the finite-temperature propagators corresponds to
inserting Matsubara frequencies into the momentum argument of the
zero-temperature propagator functions. From an RG point of view, this
represents the zeroth-order approximation to the full temperature-dependent
propagators. Nevertheless, we expect that this already provides a quantitative
insight into the finite-temperature phase structure for the following reasons:
finite-temperature modifications of the propagators are expected to occur for
momentum scales below the temperature scale and, more prominently, for
temperatures below $\Tc$, as is confirmed by corresponding functional and
lattice studies \cite{Gruter:2004bb}. By
contrast, the effective order-parameter potential is dominantly built up from
momentum modes near the scale $2\pi T$.  Therefore, detecting $\Tc$ from
above, the IR properties of the propagators are hardly probed and only the
decisive mid-momentum region together with the perturbative high-momentum tail
of the propagators effectively enters in the computation of the
potential. Neglecting a potentially strong explicit temperature dependence of
the propagators for small temperatures $T\ll T_{\rm c}$ and/or momentum modes
below the temperature scale is thus an acceptable approximation for detecting
the phase boundary. The validity of this approximation has been verified
explicitly for scalar theories in Ref.~\cite{Braun:2009si}. 

In addition to our analytical discussion of the confinement criterion, we have
now used the full functional flow equation including the term depending on
$\partial_t \Gamma^{(2)}_k$ in Eq.~\eqref{eq:iIflowpart} in our numerical
study. We observe that the order of the phase transition for a given gauge
theory remains unchanged upon the inclusion of this term. Moreover the phase
transition temperature increases only by $\lesssim 7\%$ when this term is
added. For a qualitative understanding of the order-parameter potential as
discussed in the preceding section, the omission of this term is hence
justified which confirms the picture arising from the our confinement
criterion.  For details on the numerical computation of the order-parameter
potential, we refer to App.~\ref{App:Details}.

In order to convert our results into physical units, we fix our propagators
relative to the lattice scales. In turn, the propagators on the lattice can be
converted into physical units by measuring lattice momenta in units of, e.g.,
the string tension. In this manner, we can determine $\Tc$ in physical units
corresponding to a string tension of $\sigma=440$ MeV. In our studies we 
keep the position of the peak of the gluon propagator fixed for all gauge groups.
This provides a prescription for a comparison of lattice results for $\Tc$ and 
our results.

\subsection{$SU(N)$}
\begin{figure}[t]
\includegraphics[width=1\linewidth]{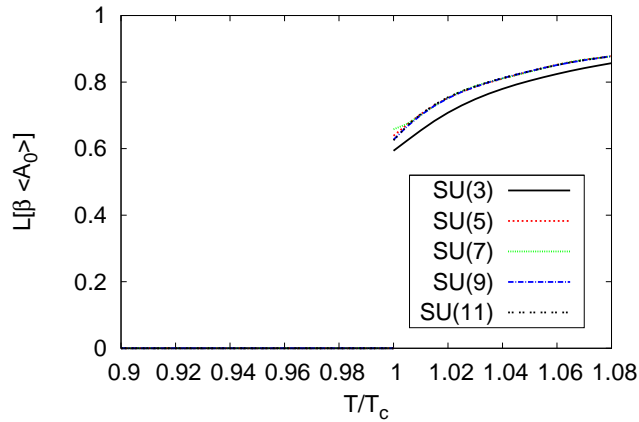}
\caption{Polyakov loop $L[\BA]$ as a function of temperature
  for SU(3), SU(5), SU(7), SU(9), SU(11). We observe that the phase transition
  is of first order. Hardly any difference for the order parameter is visible
  for $N\geq 5$, suggesting a close proximity of these gauge groups to the
  large-$N$ limit.}
\label{fig:SUN}
\end{figure}

Let us first consider the gauge groups $SU(N)$.  In Fig.~\ref{fig:SUN}, we
show our results for the order parameter $L[\BA]$ for $N=3,5,7,9,11$ as a
function of $T/\Tc$; the corresponding results for $SU(2)$ and $SU(12)$ can be
found in Figs.~\ref{fig:Sp2} and~\ref{fig:E7}, respectively. We find a second
order phase transition for $SU(2)$ and a first-order phase transition for
$SU(N)$ ($N=3,4,\dots,12$).  For $SU(2)$ the phase transition occurs at
$\Tc\approx 265\,\text{MeV}$. For $SU(3)$ we find $\Tc\approx
291\,\text{MeV}$. With increasing rank of the gauge group,
the phase transition temperature increases slightly and approaches $\Tc\approx
295\, \text{MeV}$ for $SU(5)$. For $N\geq 5$ our results for the
order-parameter are essentially independent of $N$.  In other words, our
results for the phase transition temperature for $SU(3)$ is already close to
the large-$N$ value. This independence of the Polyakov loop on $N$ for
  $N\geq 5$ is in accordance with recent lattice studies of $SU(N)$ Yang-Mills
  theories~\cite{Panero:2009tv}. Since we employ the same
propagators for all $SU(N)$, the increase of $\Tc$ is only due to the increase
in the rank of the gauge group.  In accordance with the weak dependence of
$\Tc$ on $N>2$, the order-parameter as a function of $T/\Tc$ depends only
slightly on the rank of the gauge group.  Note that our result for the
Polyakov loop $L[\BA]$ for $T/\Tc >1$ is higher than the corresponding
expectation value $\langle L[\mathsf{A}_0] \rangle$ of the Polyakov loop as
obtained from lattice simulations, being in perfect agreement with the
Jensen inequality $L[\langle \mathsf{A}_0\rangle]\geq \langle
L[\mathsf{A}_0]\rangle$.

At this point we would like to emphasize that our studies are of
course not bound to $N\leq 12$. Our approach can be straightforwardly
generalized to $N>12$ with the aid of Eq.~\eqref{eq:Vsum}. Our
numerical study of a given gauge group involves three simple steps:
computing $V_{\rm SU(2)}$, finding the eigenvalues of the generators
of the Cartan subalgebra in the adjoint representation for the gauge
group under consideration, and finally minimizing
Eq.~\eqref{eq:Vsum}. Therefore the computation of the order parameter
for very large gauge groups is not considerably more involved than for
smaller ones.

In view of the approximations listed above, we expect corrections to our
results from modifications of the propagators due to finite temperature and
due to order-parameter fluctuations.  Whereas finite-temperature corrections of
the propagators affect the results for the order parameter of all gauge
groups, order-parameter fluctuations play a particularly important role in
$SU(2)$ since it has a second-order phase transition.

\subsection{$Sp(2)$}
\begin{figure}[t]
\includegraphics[width=1\linewidth]{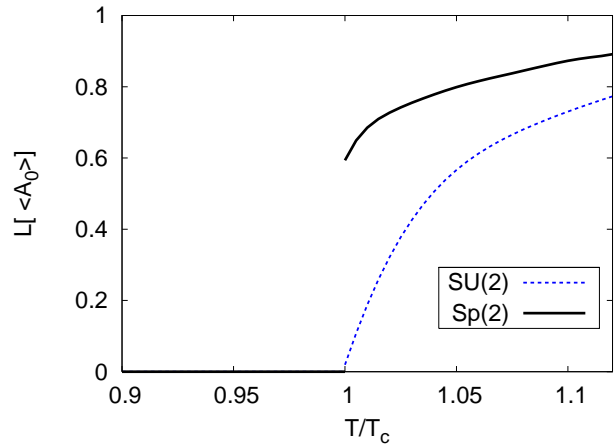}
\caption{Polyakov loop $L[\langle A_0\rangle]$ 
for $SU(2)$ (blue/dashed line) and $Sp(2)$ (black/solid line). The phase transition 
is of second order for $SU(2)$ and of first order for $Sp(2)$.}
\label{fig:Sp2}
\end{figure}

The fact that confinement and center symmetry are related naively suggests
that gauge groups with the same center may show similar phase transition
properties. This is, however, not the case as the prime counter-example of
$SU(2)$ vs. the symplectic group $Sp(2)$ demonstrates: both gauge groups have
the same center $Z(2)$, but exhibit qualitatively different phase-transition
properties\footnote{In our conventions $Sp(1)$ is isomorphic to $SU(2)$.}.  
Our results for the order parameter $L[\langle {A_0}\rangle]$ as
a function of $T/\Tc$ for $SU(2)$ and $Sp(2)$ are depicted in
Fig.~\ref{fig:Sp2}. The generators of $Sp(2)$ are given in
App.~\ref{App:Sp2Gen}.
We find a second-order phase transition for $SU(2)$
and a first-order phase transition for $Sp(2)$. Therefore, $SU(2)$ falls into
the Ising universality class~\cite{Svetitsky:1985ye} but $Sp(2)$ does
not. Moreover the phase transition temperature for $Sp(2)$ gauge theory is
close to the value of $SU(3)$ gauge theory; we find $\Tc\approx
286\,\text{MeV}$. Since we use the same propagators for both gauge groups and
the center of both groups is $Z(2)$, it is natural to relate the different
nature of the phase transition to the different dimensionality of the two
groups \cite{Holland:2003kg}. In fact, the number of degrees of freedom in the
deconfined phase is much larger in $Sp(2)$ than in $SU(2)$. This strong
mismatch in the number of dynamical degrees of freedom in the confined and
deconfined phase appears to enforce a first-order phase transition in
$Sp(2)$. In this respect these findings resemble the situation in the case of
$SU(3)$ in $2+1$ and $3+1$ space-time dimensions. While the phase transition
in $SU(3)$ is of first order in $3+1$ dimensions, it is of second order in
$2+1$ dimensions~\cite{Engels:1996dz}. Again, this might be traced back to the
fact that the mismatch in the number of dynamical degrees of freedom in the
deconfined phase is smaller in $d=2+1$ than it is in $d=3+1$.  

Our findings for the nature of the $Sp(2)$ phase transition are in
accordance with lattice
simulations~\cite{Pepe:2006er,Holland:2003kg}. Quantitatively, our
approximation of neglecting terms $\propto V''(\BA)$ on the right-hand
of Eq.~\eqref{eq:iIflowpart}, which account for order-parameter
fluctuations, might be more severe in $Sp(2)$ due to its similarity to
$SU(2)$. The inclusion of these fluctuations may only lead to a
  weaker first-order jump and an increase of the critical
  temperature.

\subsection{$E(7)$}
\begin{figure}[t]
\includegraphics[width=1\linewidth]{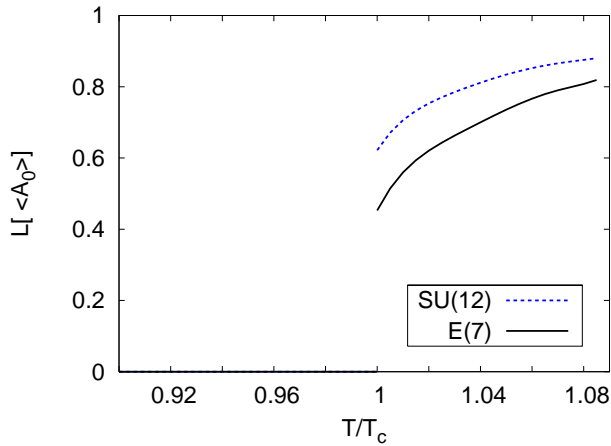}
\caption{Polyakov loop $L[\BA]$ for $SU(12)$ (blue/dashed line) and $E(7)$
  (black/solid line).  The phase transition is of first order for both
  $SU(12)$ and $E(7)$ gauge theory.}
\label{fig:E7}
\end{figure}

Another interesting test of the proposal that the order of the phase
transition is related to the size of the gauge group \cite{Holland:2003kg} is
the following comparison between $SU(12)$ and $E(7)$ gauge theory.  The
dimension of the adjoint representation of these two gauge groups is about the
same: We have $d_{\rm adj}=133$ for $E(7)$ and $d_{\rm adj}=143$ for $SU(12)$.
The gauge groups differ with respect to their center, being $Z(2)$ for $E(7)$
and $Z(12)$ for $SU(12)$.

In Fig.~\ref{fig:E7}, our result for the Polyakov loop $L[\BA]$ for both
theories is depicted as a function of $T/\Tc$.  As discussed above,
the phase transition in $SU(12)$ is of first order and occurs at $\Tc\approx
295\,\text{MeV}$. For $E(7)$, our RG approach predicts a first-order phase
transition at $\Tc\approx 295\,\text{MeV}$ as well.  

Our study is thus compatible with the suggestive relation of the order of the
phase transition and the mismatch of the number of degrees of freedom above
and below the phase transition -- provided the glueball spectrum below the
phase transition in $E(7)$ is similar to that of $SU(N)$.

However, we also observe that the height of the jump of the order
parameter is smaller in $E(7)$ than in $SU(N)$ for all values of $N$
studied in the present paper. Even though the Polyakov loop $L[\BA]$
is not an RG invariant quantity, our approach of studying the
associated eigenvalue distribution allows us to give the height of the
jump a physical meaning. This suggests that the mismatch in the number
of degrees of freedom is not the only mechanism that determines the
nature of the phase transition.

\begin{figure}[t]
\includegraphics[width=1\linewidth]{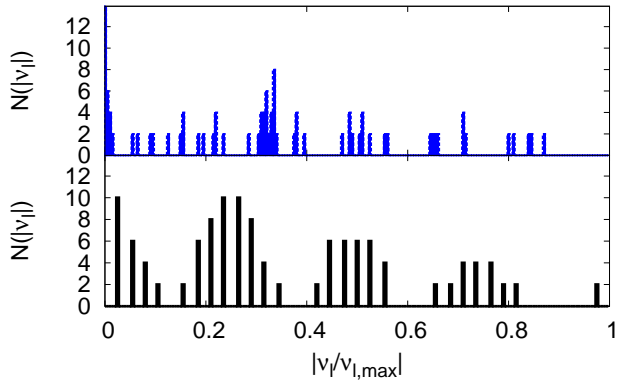}
\caption{Eigenvalue distribution $N(|\nul|)$ of the spectrum \eqref{eq:nul}
as a function of the (normalized) eigenvalue $|\nul/\nul{}_{\rm max}|$ for $E(7)$ 
(lower panel) and $SU(12)$ (upper panel) at the ground state of the potential 
for $T\to \Tc^+$.}
\label{fig:eigenvalues}
\end{figure}

In order to gain a better understanding of the nature of the phase transition,
we study the eigenvalue distribution $N(|\nul|)$ of the spectrum of the color
matrix in the Cartan subgroup as defined in \Eqref{eq:nul}.  In
Fig.~\ref{fig:eigenvalues} we show $N(|\nul|)$ as a function of the normalized
eigenvalues $|\nul/\nul{}_{\rm max}|$ at the ground state of the order-parameter potential for
$T\to \Tc^+$, approaching the critical temperature from above.  Here, the 
eigenvalues have been binned with a bin size of $\Delta
\nul = 0.005$.

The eigenvalues correspond to Fourier frequencies of SU(2) Weiss potentials,
cf. \Eqref{eq:Weiss} and \Eqref{eq:Vsum}. For instance, the higher the dominating Fourier
frequency in the high-temperature phase, the closer the $\phi$ minimum is to
$\phi=0$, implying that $L[\BA]$ is closer to $L[\BA]=1$.  By contrast, if
lower eigenvalues dominate, $L[\BA]$ can approach the center ordered state in
a smoother fashion. This is precisely what we observe for $E(7)$ in
contradistinction to the $SU(N>2)$ gauge groups, where the eigenvalues cluster
around $\nul/\nu_{l{\rm max}} \approx 0, 0.25, 0.5, 0.75$, leading to
an almost constructive interference of $SU(2)$ potentials with almost
identical periodicity.  We stress that the eigenvalue distribution
$N(|\nul|)$ depends on the actual position $\BA _{\rm min}$ of the
ground-state of the potential. Therefore the eigenvalue distribution
and hence the strength of the first-order phase transition depends on
the actual trajectory $\BA _{\rm min}(T)$ of the physical ground-state
close to $\Tc$ in the space spanned by the generators of the Cartan
subalgebra. Of course, since $L[\BA]$ provides only an upper bound for
$\langle L \rangle$ it is not immediately clear whether the difference
in $L[\BA]$ which we observe for $E(7)$ and $SU(N)$ also translates
into a similar difference in $\langle L\rangle$. If so, we expect the
phase transition for $E(7)$ to be smoother than for $SU(N)$. Taking
into account that order parameter fluctuations dropped so far can
smoothen the phase transition even further, our results may not even
be taken as a strict excluding evidence for a second order phase
transition in $E(7)$.

With respect to our study of $Sp(2)$ we indeed find that the eigenvalue
distribution $N(|\nul|)$ exhibits a pattern very similar to the one of $SU(3)$
resulting in a jump of the order parameter at the phase transition with a
height comparable to the one of $SU(3)$ Yang-Mills theory, see
Figs.~\ref{fig:SUN} and~\ref{fig:Sp2}.

\vspace{-.1cm}
\section{Conclusions}
In the present paper we have discussed the nature of the phase
transition in various gauge groups and based on a simple confinement
criterion put forward in Ref.~\cite{Braun:2007bx}. For our study of
the critical dynamics of non-abelian gauge theories we computed the
order-parameter potential in Landau-DeWitt gauge by employing gauge
correlation functions. In particular, for the question whether the
system is in the confining phase near zero temperature, the
quantitative knowledge of zero-temperature propagators is
sufficient. Even for questions related to the nature and quantitative
properties of the phase transition near the critical temperature, we
have argued that we expect that zero-temperature propagators provide
for a reasonable approximation. Of course, the inclusion of knowledge
about thermal propagators in our formalism is straightforwardly
possible. Moreover we have dropped order-parameter fluctuations which,
though irrelevant for the question of confinement near zero
temperature, we expect to affect, for instance the quality of, e. g.,
critical exponents. The fluctuations have been included in the
$SU(2)$-study in Polyakov gauge, \cite{Marhauser:2008fz}, and lead to
Ising-class critical exponents as expected.

In agreement with lattice simulations we have found a first-order phase
transition for $SU(N)$ gauge theories with $N\geq 3$. For $SU(2)$ the phase
transition is of second order. Moreover we have studied $Sp(2)$ and $E(7)$
gauge theory and compared our results to $SU(N)$ Yang-Mills theory.  In
agreement with lattice simulations~\cite{Pepe:2006er,Holland:2003kg} we
observe a first-order phase transition in $Sp(2)$. As a new set of
characteristic quantities of the phase transition, we have introduced the
distribution of eigenvalues of $\BA$ which within the Polyakov gauge can be
related to an order parameter. These eigenvalues serve as Fourier frequencies
of a superposition of $SU(2)$ Weiss potentials yielding the full
nonperturbative Weiss potential $V[\BA]$ from which $L[\BA]$ can be deduced 
as an upper bound to the Polyakov loop order parameter. 
Whereas the order-disorder nature of the deconfinement transition is related
to center symmetry, the center degrees of freedom themselves are not always
the relevant degrees of freedom to understand the phase transition. 
For instance, this is obvious in the case of $Sp(2)$ which has a center $Z(2)$ but
exhibits a phase transition different from $SU(2)$ with the same
center. This is in fact illustrated by the eigenvalue distribution: for 
$Sp(2)$ the eigenvalue distribution at the phase transition is
similar to the one of $SU(3)$ as are other properties of the phase transition.
In this picture the 
order-parameter potential can be considered as a destructive
interference/superposition of $SU(2)$ potentials favoring a first-order phase
transition. Moreover, we have a stronger mismatch in the number of the
dynamical degrees of freedom in $Sp(2)$ in the confined and deconfined phase
compared to $SU(2)$ Yang-Mills theory~\cite{Pepe:2006er,Holland:2003kg}.

For $E(7)$ gauge theory we find that the phase transition is of first order as
well.  Here, the mismatch in the number of dynamical degrees of freedom in the
confined and deconfined phase is even stronger than it is in $Sp(2)$,
suggesting that the first-order phase transition is even stronger.  However,
our RG study suggests that the first-order phase transition is weaker for
$E(7)$ than it is for $SU(12)$ or $Sp(2)$. We have argued that this weak
first-order transition can be traced back to the eigenvalue distribution at
the phase transition. In contrast to $Sp(2)$ and $SU(N)$ we have found that
the distribution exhibits distinct equidistant maxima resulting in an almost
constructive interference of $SU(2)$ potentials. In this respect $E(7)$ is
closer to $SU(2)$ than to $SU(N)$ with $N\geq 3$. However, further studies are
needed to establish this picture. For example, a RG study of $SU(3)$ and
$Sp(2)$ Yang-Mills theory in $2+1$ dimensions may help to shed more light on
the underlying mechanisms of the deconfinement phase transition since it is
known from lattice simulations that the nature of the phase transition in both
gauge groups changes from first to second order when the number of dimensions
is reduced~\cite{Engels:1996dz,Pepe:2006er,Holland:2003kg}.  

Another interesting case is the gauge group $G(2)$ with non-trivial
center.  In this case it has been
found~\cite{Pepe:2006er,Wellegehausen:2009rq} that the Polyakov loop
exhibits a jump but is non-vanishing for all temperatures.  A
verification of our quark confinement criterion with the aid of G(2)
Yang-Mills theory is under way and will help us to establish our
findings and to improve our understanding of confinement in gauge
theories.

\acknowledgments Helpful discussions with A.~Wipf on $E(7)$ gauge
theory are gratefully acknowledged.  This work was supported by the
DFG support under Gi 328/1-4 and Gi 328/5-1 (Heisenberg program) and
through the DFG-Research Training Group "Quantum- and Gravitational
Fields" (GRK 1523/1). JMP acknowledges support by Helmholtz
Alliance HA216/EMMI.

\appendix

\section{Details on the computation of the order-parameter potential}\label{App:Details}
\begin{figure}[t]
\includegraphics[width=1\linewidth]{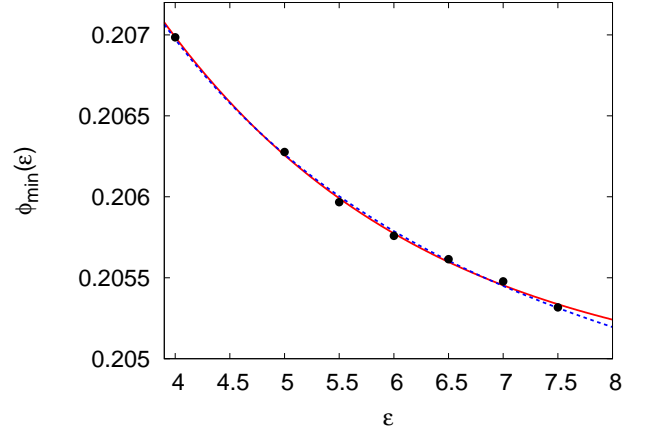}
\caption{$\phi_{\rm min}$ as function of the 'smearing' parameter $\epsilon$ for $T=300\,\text{MeV}$. 
The numerical data is depicted by dots. The red (solid) and blue (dashed) line are the results 
from the fit to the functions given in the first and second line of Eq.~\eqref{eq:fitfct}, 
respectively.}
\label{fig:epsdep}
\end{figure}

In this addendum we discuss some details on the computation of the 
order-parameter potential. The order-parameter potential can be obtained directly 
from an evaluation of Eq.~\eqref{eq:iIflowpart}. While the first term in Eq.~\eqref{eq:iIflowpart}
is independent of our choice of the regulator function, the second term is not.
Since we employed the optimized regulator~\eqref{eq:optprop}, we encounter expressions 
involving unit-step functions in the second term on the right-hand side of 
Eq.~\eqref{eq:iIflowpart}. These unit-step functions depend on the background field $\BA$. 
This dependence on $\BA$ generates divergent RG
flows for $\Lambda\to\infty$. In principle one can deal with these divergences by 
computing appropriate counter-terms at the initial UV scale. In the present paper 
we sought for a different approach to circumvent this problem and 
introduced 'smeared' unit-step functions:
\begin{eqnarray}
f_{\theta} (x[\BA],\epsilon)&=&{\rm e}^{-(x[\BA])^{\epsilon}},\nonumber\\ \label{eq:regeps}
\lim _{\epsilon\to\infty} f_{\theta} (x[\BA],\epsilon)&=&\theta(1-x[\BA])\,,
\end{eqnarray}
where $x[\BA]$ is an arbitrary function depending on the background field $\BA$.
Using $f_{\theta} (x[\BA],\epsilon)$ instead of $\theta(1-x[\BA])$ 
yields an order-parameter potential periodic in $\BA$ for any finite value of $\epsilon$
and allows to get conveniently rid of the unphysical divergent parts of the flow. 
For our numerical study of the deconfinement phase transition we
have used $\epsilon=7$. In Fig.~\ref{fig:epsdep} we illustrate the dependence of the position
of the minimum $\phi_{\rm min}^{\rm fit}=\beta\BA_{\rm min}/(2\pi)$ 
of the potential on the 'smearing' parameter $\epsilon$ for $T=300\,\text{MeV}$ for $SU(2)$ 
Yang-Mills theory. From an extrapolation of our results to $\epsilon\to\infty$ using the two functions
\begin{eqnarray}
\phi_{\rm min}^{\rm fit}(\epsilon)=
\begin{cases}
    {\rm const.} + a\,{\rm e}^{-b\epsilon}\\
    {\rm const.} + \frac{c}{\epsilon}
  \end{cases}\,,\label{eq:fitfct}
\end{eqnarray}
where $a$, $b$ and $c$ are fit parameters, we estimate that the theoretical error is less 
than $1\%$ when $\epsilon=7$ is used. Note that the fit function in the second line
of Eq.~\eqref{eq:fitfct} can be formally deduced from a Taylor expansion  around $\epsilon$ 
of the integral of a general polynomial in $x[\BA]$ weighted by $f_{\theta} (x[\BA],\epsilon)$.

\section{Generators of $Sp(2)$}\label{App:Sp2Gen}
Our definition of the generators of $Sp(2)$ in the fundamental representation is
as follows:

\begin{eqnarray}
 C_1&=&\left(
\begin{array}{cccc}
 0 & i & 0 & 0 \\
 i & 0 & 0 & 0 \\
 0 & 0 & 0 & 0 \\
 0 & 0 & 0 & 0
\end{array}
\right),\nn
\end{eqnarray}

\begin{eqnarray}
C_2&=&\left(
\begin{array}{cccc}
 0 & 1 & 0 & 0 \\
 -1 & 0 & 0 & 0 \\
 0 & 0 & 0 & 0 \\
0 & 0 & 0 & 0
\end{array}
\right),\nn
\end{eqnarray}

\begin{eqnarray}
C_3&=&\left(
\begin{array}{cccc}
 i & 0 & 0 & 0 \\
 0 & -i & 0 & 0 \\
 0 & 0 & 0 & 0 \\
 0 & 0 & 0 & 0
\end{array}
\right),\nn
\end{eqnarray}

\begin{eqnarray}
C_4&=&\left(
\begin{array}{cccc}
 0 & 0 & 0 & 0 \\
 0 & 0 & 0 & 0 \\
 0 & 0 & 0 & i \\
 0 & 0 & i & 0
\end{array}
\right),\nn
\end{eqnarray}

\begin{eqnarray}
C_5&=&\left(
\begin{array}{cccc}
 0 & 0 & 0 & 0 \\
 0 & 0 & 0 & 0 \\
 0 & 0 & 0 & 1 \\
 0 & 0 & -1 & 0
\end{array}
\right),\nn
\end{eqnarray}

\begin{eqnarray}
C_6&=&\left(
\begin{array}{cccc}
 0 & 0 & 0 & 0 \\
 0 & 0 & 0 & 0 \\
 0 & 0 & i & 0 \\
 0 & 0 & 0 & -i
\end{array}
\right),\nn
\end{eqnarray}

\begin{eqnarray}
C_7&=&\left(
\begin{array}{cccc}
 0 & 0 & \frac{i}{\sqrt{2}} & 0 \\
 0 & 0 & 0 & -\frac{i}{\sqrt{2}} \\
 \frac{i}{\sqrt{2}} & 0 & 0 & 0 \\
 0 & -\frac{i}{\sqrt{2}} & 0 & 0
\end{array}
\right),\nn
\end{eqnarray}

\begin{eqnarray}
C_8&=&\left(
\begin{array}{cccc}
 0 & 0 & \frac{1}{\sqrt{2}} & 0 \\
 0 & 0 & 0 & \frac{1}{\sqrt{2}} \\
 -\frac{1}{\sqrt{2}} & 0 & 0 & 0 \\
 0 & -\frac{1}{\sqrt{2}} & 0 & 0
\end{array}
\right),\nn
\end{eqnarray}

\begin{eqnarray}
C_9&=&\left(
\begin{array}{cccc}
 0 & 0 & 0 & \frac{i}{\sqrt{2}} \\
 0 & 0 & \frac{i}{\sqrt{2}} & 0 \\
 0 & \frac{i}{\sqrt{2}} & 0 & 0 \\
 \frac{i}{\sqrt{2}} & 0 & 0 & 0
\end{array}
\right),\nn
\end{eqnarray}

\begin{eqnarray}
C_{10}&=&\left(
\begin{array}{cccc}
 0 & 0 & 0 & \frac{1}{\sqrt{2}} \\
 0 & 0 & -\frac{1}{\sqrt{2}} & 0 \\
 0 & \frac{1}{\sqrt{2}} & 0 & 0 \\
 -\frac{1}{\sqrt{2}} & 0 & 0 & 0
\nn
\end{array}
\right).
\end{eqnarray}

\end{document}